\documentclass[aps,prl,twocolumn,10 pt,showpacs]{revtex4}
\usepackage{amssymb}

\usepackage{graphicx}

\begin{document}

\title{Experimental study of the correlation length of critical-current
    fluctuations in the presence of surface disorder: Probing vortex
    long-range interactions.}
\author{J. Scola, A. Pautrat, C. Goupil, Ch. Simon}
\affiliation{CRISMAT, UMR 6508 du CNRS et de l'ENSI-Caen, 6 Bd
Mar\'echal Juin, 14050 Caen, France.}

\begin{abstract}
We report on critical currents and voltage noise measurements in
Niobium strips in the superconducting state, in the presence of a
bulk vortex lattice ($B < B_{C2}$) and in the surface
superconducting state ($B_{c2}< B < B_{C3}$). For homogeneous
surfaces, the correlation length of the current fluctuations can
be associated with the electromagnetic skin depth of vortex
superficial instabilities. The modification of the surface state
by means of low energy irradiation induces a strong modification
of the critical current and of the noise. The appearance of a
corner frequency in the spectral domain can be linked with the low
wave-vectors of the artificial corrugation. Since this latter
occurs only for $B < B_{C2}$, we propose that the long-range
interactions
 allow the correlation length to extend up to values imposed by the surface topography.

\end{abstract}


\newpage
\maketitle
 Noise measurements are powerful tools to go inside the origin
 of the vortex pinning, and of the dynamical interactions between the vortex lattice and the sample disorder. This noise, generated during the lattice flow, is called $\textit{the flux-flow noise}$ \cite{clem}. It is generally characterized by the shape of its spectral density,
   and by its power. As it is often proposed in noise analysis, it is convenient to define a correlation length,
 within which the fluctuations are correlated. This defines the fluctuator of the system.
  To the extent that the fluctuators are independent and that the system is large enough, the
 correlation length can be calculated from
 the noise power via an usual statistical averaging (the central limit
 theorem) \cite{mike1}. Recently, it was shown that no difference can be observed between $\textit{the flux-flow noise}$ in the mixed state and in the surface
 superconducting state of a Niobium slab \cite{jo2}. In other words, the same
 fluctuator is present with or without a bulk vortex lattice, showing clearly its superficial origin. The coupling to the bulk was shown to be
due to the conservation of the total current.  This confirms previous auto and cross correlation experiments of both flux and voltage noises \cite{bernard}, and explains the insensivity of the low frequency noise to bulk perturbations \cite{jo1}, in Pb-In alloys.
 In these experiments, two parameters,
 the normalized spectrum of the fluctuator and its correlation length, are experimentally justified but are not explained \cite{bernard,jo2}. If the fluctuator is of superficial
 origin, it should be possible
 to induce notable changes in the noise characteristics after some surface treatments.
 A following change in the noise spectral density, the noise power, or eventually in the underlying statistics, would give some clues to understand the spectrum and the fluctuator origins.
 Wherever it takes
place, it has been made clear for years that the existence
 of flux-flow noise is intimately linked to the nature of the
 relevant disorder, i.e. to the vortex pinning. In the conclusion of his review article, Clem states that such a \textit{flux-flow noise} theory
 `` should be intimately related to an appropriate theory of critical current density ''  \cite{clem}. This implies that any noise analysis would be notably improved
 if some characterization of the vortex pinning is proposed beforehand.

  In this paper, we are thus interested in the modification of $\textit{the flux-flow noise}$ by a tuning of the relevant
 disorder.
 For that, we will first discuss the nature of the pinning in our
 samples and bring some experimental arguments on the way that the
 critical current can be substantially modified by a surface treatment. In a second
approach, the noise mechanism proposed in previous papers \cite{bernard,jo1}, will be discussed and specified by some recent results concerning the
physical origin of the fluctuator. In the last part of
 the paper, the effect of surface irradiation on the noise power and the noise spectrum will be
 shown and discussed. For sake of generality, these
 measurements are made in Niobium, a well documented and conventional type II
 superconductor.

\section{Experimental}

All measured samples are parts of the same piece of bulk Nb
(initially 2 $mm$ thick and 12 $mm$ long). The thickness $t$ was
first roughly reduced to 0.23 $mm$, and the surfaces were
progressively polished with 27 $\mu m$, 15 $\mu m$ then 7 $\mu m$
rough papers. Then, three samples of different widths $W$ were cut
by a wire. However, the roughness obtained after the finest
mechanical polishing was still too large to reduce sufficiently
the critical current. In order to reduce efficiently this surface
roughness, and to eliminate the $Nb_{2}O_{5}$ oxide sheath which
naturally develops on the niobium surface, a buffer chemical
polishing (BCP) was performed. Each sample was plunged into an
equivolume solution of $H_3PO_4 (85\%), HF (40\%), HNO_3 (69\%)$
during eight minutes. Previous reports of the rms roughness after such a
BCP give values close to 1 $nm$ in the $\mu$m scale \cite{rugosite}, in agreement with our AFM
measurements. From the analysis of the AFM pictures (see fig. 2),
we find 0.7 $nm$ rms. This value is the mean values of 10 different line scans taken in an
area of 10$\times$10 $\mu$m$^{2}$. After this step, all samples have similar
surface state and their bulk properties are unchanged. In the
following, the samples (1) and (2) will correspond to the samples
with respectively $W$=0.25 $mm$ and $W$= 1.24 $mm$ ($t$= 0.22 $mm$
for each). The sample (3) will refer to the irradiated sample.
Each large face of this sample was irradiated by an argon beam for
30 minutes (acceleration voltage = 600 $V$, and argon pressure
$\approx$ 2.10$^{-4}$ $mbar$). The sample holder was inclined at
45 degrees and continuously rotated in order to make the etching
very uniform in a 10 $mm$ $\times$ 10 $mm$ area. Such low energy
irradiation causes only superficial damages (about 10 $nm$ of
depth), which will be discussed in the text.

 The superconducting parameters of each samples were measured by DC transport, magnetization (SQUID) and specific heat measurements.
The following parameters were measured: $T_{c} =$ 9.2 $\pm$ 0.1
$K$, $B_{c2}(4.2 K) =$ 2900 $\pm$ 50 $G$ and $\rho_n (10 K)
\approx$ 0.5 $\mu\Omega.cm$. Comparing with the data of the ref.
\cite{niobium}, a Ginzburg-Landau parameter $\kappa=$ 0.84 can be
inferred. The critical currents have been measured by
voltage-current characteristics.

In addition, we have measured the voltage noise induced during the
vortex lattice motion. Voltage signals were recorded with the four
probes method, amplified by ultra low noise preamplifier
($SA-400F$, 0.7 $nv/ \sqrt{Hz}$). This signal source is then
plugged into the analog inputs of a dynamic signal analyzer
($PCI-4551$), converted into digital signals and mathematically
processed. All the experimental set-up was electromagnetically
shielded to avoid much of the external disruption. Such noise
measurements are extremely sensitive to the temperature stability:
a special care has been taken to avoid any excess of heating
responsible of low frequency noise (contact noise or flicker noise
due to the Helium boiling). Since Niobium is a very good conductor
($R \lesssim$ 10 $\mu \Omega$ for our samples), the contacts
resistances are the major limiting problem. Making good electrical
contacts on the bulk Niobium samples is difficult. The best
solution was to mechanically strip the end of the slabs of its
oxide sheath, to deposit a thin metallic layer and to press
between small copper pieces. Finally, this was good enough to
apply $I \lesssim$ 8 $A$ with no spurious noise. But in practice,
in our macroscopic samples, this restricts the experimental region
relatively close to $B_{c2}$.

\section{Results and discussion}

\subsection{Critical currents and vortex pinning: width and surface roughness effects}

In fig.1 are shown the critical currents versus the magnetic field
close to $B_{c2}$ for the Niobium samples. Recent works have
attributed much of the irreversible properties of type II
superconductors, including the vortex noise, to the edges of the
samples (geometrical barriers or Bean-Livingston barriers)
\cite{zeldov}. In order to quantify this effect in our samples, we
have measured the critical current of the samples (1) and (2),
which have a very contrasted width. If the critical current $I_c$
is principally concentrated near the sample edges, its value
should not significantly depend on the sample width. In contrast,
since we measure a simple linear relation between $I_{c}$ and W (fig.
1), $I_c$ can be considered as macroscopically
homogeneous all over the width of the surface. Note that,
$\textit{a priori}$, it does not exclude small inhomogeneities and
local $I_c$ variations near the lateral edges, but those effects
turn out to be on average negligible in our samples. We will show
hereafter that the noise measurements will be a more precise
probe. Finally, macroscopic homogeneity of the critical current
means that a critical current per unit of width $i_c$ (A/m) =
$I_c$/(2$W$) is a relevant description of the critical properties.

For conventional soft type II superconductor, except the
irreversible edges effects, the vortex pinning has been shown to
arise from pinning of the tips of the vortices on the top and
bottom surfaces. This is quantified by the following expression
(\cite{fib} and references herein):

\begin{equation}
I_c / 2 W =  i_{c} (A/m) = \varepsilon.sin(\theta_c)
\end{equation}

where $\varepsilon$ stands for the overall equilibrium
magnetization, and $\theta_c$ is the characteristic surface
roughness angle.

The topography of the surfaces was analyzed using atomic force
microscopy (AFM) in the tapping mode. The roughness of the virgin
surfaces exhibits a random disorder, with no characteristic scale.
The fig.2 is a 3D representation of the surface roughness. In
terms of vortex surface pinning potential, the relevant parameter
is equivalent to a contact angle $\theta$. In order to calculate
it, we first take a statistical representative cross section
$h(x)$ (the average of several cross sections), then calculate its
derivative $dh/dx$ = tan ($\theta$). The spectral density is then
$S_{\theta\theta} = FFT( \langle \theta(x) \theta(x+x')\rangle)$.
The main roughness angle accessible for the vortices is given by
\cite{fib}:

\begin{equation}
\overline{\theta^{2}}=\int_{K_{min}}^{K_{max}} S_{\theta\theta}
dK\approx\int_{0}^{2 \pi / \xi} S_{\theta\theta} dK
\end{equation}

where $\xi = \sqrt{\phi_ 0/ 2\pi B_{c2}}\approx$ 33 $nm$ is the
coherence length.

In order to calculate the critical current with the equation (1),
the overall equilibrium magnetization should be known. Close to
$B_{c2}$, $\mu_0 \varepsilon = \frac{B-B_{c2}}{\beta (2 \kappa^2 -
1)}$ in the Abrikosov limit, with $\beta$ = 1.16 for the hexagonal
symmetry of the vortex lattice. Using $B_{c2} =$ 0.29 $T$ and
$\kappa$ = 0.84 in the equation (1), one finds, at $B$ = 0.25 $T$,
$\theta_c = \arcsin (i_c/\varepsilon)$ = 2.1 $\pm$ 0.1 $deg$. The
surface topography analysis of the samples (1) leads to $\theta$ =
2.3 $\pm$ 0.2 $deg$ (equation (2)), in a close agreement. The
equation (1) appears thus to describe the critical current data,
with $\theta \approx \theta_c$ as a reasonable average parameter
describing the surface pinning potential. We note that the same
analysis was successfully applied in Niobium films \cite{fib}.
This confirms the previous assumption \cite{fib} that the
difference of $J_c(A/m^2)=$ 2 $i_c/t$ between bulk and thin films
Nb (with different thickness $t$) does not mean a stronger pinning
in films. This simply proves that $J_c$ is not the good parameter,
but that $i_c$ is.

We now focus on the consequence of the surface irradiation on the
vortex pinning. The measured critical current has been increased by a factor $\times$ 2.9 in the sample (3) (fig.1).
 As evidenced in the fig.2, the irradiation results in a strong
 degradation of the surface, with both an increase of the overall roughness and the appearance of large craters. Using the surface analysis and the equation (2), $\theta =$ 4 $\pm$ 0.2 $deg$ is deduced
 (fig.3). The roughness angle has thus been increased by a factor $\times$ 1.7 by the
 irradiation. Note that the important parameter is not directly the roughness but the roughness angle which is a derivative. As a consequence, the increase of this latter is not as strong as a visual impression could give (see fig. 2).
 Finally, the increase of the critical current is qualitatively explained,
 but some difference appears in the comparison of the critical
 angles. In fact, $\varepsilon$ in the equation (1) stands strictly for the surface
 value, whereas we use the only known bulk parameters for the calculation of the Abrikosov expression. If this approximation seems well justified in
 the case of homogeneous samples, it becomes questionable when the
 surfaces have been irradiated, because the concentration of
 impurities near the surfaces should locally modify the
 thermodynamic parameters. In general, the most important change is in the coherence length which has to be replaced by $\sqrt{\xi \ell} < \xi$ in the dirty limit, with $\ell$ the mean free path (see for example \cite{ulmaier}). This leads to a local increase of $B_{c2}$, here restricted close to the surfaces.
 We will show below that the noise measurements confirm this
 slight increase, providing an attractive explanation of the underestimation of the critical current after the irradiation in the mixed state.

\subsection{On the noise for the smooth surfaces}

When the applied current is over-critical, the voltage in the
mixed state is given by $V=R_{ff} (I-I_c)$, with $R_{ff}$ the
flux-flow resistance. In the surface superconducting state, the
same expression applies with $R_n$ instead of $R_{ff}$.
 It has been shown that, in the case of Pb-In alloys, most of the voltage noise
  originates from current fluctuations \cite{bernard,jo1}. Following here the same arguments, one can write $\delta V \approx R _{ff} \delta I_c$, and with the assumption of $N$ independent fluctuators under stationary conditions:
\begin{equation}
   \delta V = R _{ff} I_c / \sqrt{N} = R _{ff} I_c \sqrt{C^2/S},
\end{equation}
    where $S$ is the sample surface between the voltage pads, and $C$ the fluctuations correlation length. In the case of
  stationary fluctuations, $\delta V$ and thus $C$ have a well
  defined magnitude. In ref \cite{bernard} and \cite{jo1}, $C
  \lesssim 1 \mu m$ was found in Pb-In alloys.

   The metallic and flux-flow resistivities of pure Niobium are quite low. As a
consequence of the equation (3) and for identical geometrical
parameters, the flux flow noise $\delta V$
   should be
  largely reduced compared to the Pb-In case (or to any case of a relatively high resistance sample). This is consistent with our measurements. In addition, using the equation (3) for Pb-In or Nb, very similar $C$ values are extracted. This means that the same noise
   mechanism controls the fluctuator size. A strong confirmation that $C$ is a relevant parameter can be made when comparing two samples with
   contrasted widths $W_2 > W_1$ $\gg$ $C$ (respectively samples (2) and (1). For the same surface state and the same (B,T) conditions, we have already verified that $i_c$ is constant (no edges effects).
   If $C$ is constant too, the number of fluctuators should vary as $W$ and $\delta V \propto W^{-1/2}$. This can be observed from the variation of the number of fluctuators as a function of $W$ in
 the fig.4. This confirms that the noise arises
from the statistical averaging of small and independent noisy
domains.
   Consistently, we do not observe any cut-off in the spectral shape meaning that no size effect is
   introduced.
  In addition, the spectrum is very similar to the one observed in Pb-In alloys (fig.5).

   Since the number of fluctuators is found to be proportional to the
   surface of the samples, on can also conclude that the (low frequency broad band) noise does
   not arise from
   instabilities close to the edges of the sample. This
   contrasts to the conclusion of the ref. \cite{paltiel}, where
   the voltage noise was attributed to the dynamical annealing of a
   disordered vortex phase nucleated close to the edges. We stress
   that this latter interpretation was proposed for the peculiar case of
   the peak effect in $NbSe_2$, where two macroscopic critical
   current states coexist in the sample (\cite{marchevski,neutrons} and references herein), and where the large noise values are coming from the kinetic
    between these two states. This was shown to correspond to
    non-Gaussian noise by a second spectrum analysis \cite{mike}. No such peculiar features are observed in our
    samples where the noise is observed to be Gaussian and has been verified stationary using the same high order statistics method.
    Note that the fact that we observe stationary and Gaussian
    noise reinforces the modeling of the noise by independent
    fluctuators, and hence, is consistent with the fig.4.

We define a fluctuator by its correlation length $C$. Up to now, $C$ has been taken as an adjustable
parameter, whose relevance has been experimentally controlled (fig.4), but
whose physical basis have not been clarified. Thus, it is
necessary to go further inside the fluctuations mechanism. Placais
et al \cite{bernard} proposed that the instabilities in the vortex
lattice flow originate from the hanging and release of the
vortices on the surface defects, the hanging condition being the
local boundary condition for a vortex line. Since the same
boundary condition is the basis of a surface critical state
\cite{pat}, large fluctuations of magnitude $I_c$
   are expected. The more a vortex is bent, the more non dissipative current can flow. Locally, when a vortex leaves its hanging condition, $I_c$ decreases very
   fast and at the same time, the amount of normal current $I_c^{*}$ localized near the surface should increase due to the conservation of the total current. Thus, an electric field $E^{*} =
   \rho_{ff} J^{*} \approx \rho_{ff} J_{c} \approx \rho_{ff} 2i_{c}/\lambda_v$ is generated, $\lambda_v$ being the characteristic length scale associated to the decay of
   surface currents ($\lambda_v \approx \xi / \sqrt{2}$ close to $B_{c2}$ \cite{pat}). This is the elementary instability.

    Its duration can be estimated as follows. Using the Josephson relation, the line velocity reached by one vortex ending during the instability is given by
   $V_L^{*} \cong E^{*} / B$. To the extent that one instability occurs per vortex period (individual process), the associated diffusion time is $\tau$ = $a_0$ / $V_L^{*}$ = $a_0 B \lambda_v$ / $\rho_{ff} 2 i_c$.
   One the other hand, it is known from classical electromagnetism that, for a given time duration, the diffusion is restricted in a skin depth $\delta =  (\rho_{ff} \tau$ / $\mu_0 \pi)^{1/2}$.  This leads to

\begin{equation}
\delta \approx  (a_0 B \lambda_v / \mu_0 \pi 2 i_c)^{1/2}
\end{equation}

     For typical values ($B = 0.27 T$, $a_0 \approx 0.08 \mu m$, , $\rho_{ff} \approx \rho_n B/B_{c2} \approx 0.3 \mu\Omega.cm$, $i_c
   \approx 700$ $A/m$, $\lambda_v \approx 23 nm$), $\tau \lesssim $ 0.07 $ns$. One can note that these instabilities are extremely short compared to the flux-flow period $T_0 = a_0 / V_L \approx a_0 B t /  \rho_{ff} 2 i_c \approx$ 1 $\mu$s.
   The skin depth calculated from (4) is $\delta \approx$ 0.3 $\mu m$, a value
  very close to the correlation length $C$.  The comparison between the equation (4) and the experimental data for the Nb (1) and
  (2)
   is shown in the fig. 6 for $B < B_{c2}$. The agreement is satisfactory. For $B > B_{c2}$, in the surface superconducting state, the equation (4) is not very accurate
because $\lambda_v$ has not been estimated in this case. Nevertheless, it
appears clearly from the experimental data that there is no change
of regime. This can be explained by the fact that in Niobium, $\lambda_v \approx \xi$, which is the order of magnitude of the surface sheath \cite{fink}.
 It can be noted that in the equation (4), the
   important parameter which determines the range of the fluctuations is not
   the flux-flow resistivity but the surface critical current $i_c$.
   Nb and Pb-In have very different resistivity but similar $i_c$ values ($i_c (T=$4.2 $K$, $B/B_{c2}=$
   0.9)$\approx$ 700 $A/m$ for Nb and 300 $A/m$ for Pb-In).
   This explains why Nb and Pb-In samples exhibit also similar $C$
   values.

   One concludes that $C$ can be described as the skin length
   of very fast instabilities close to the surface.
   We stress that this mechanism determines the noise power via the size of a coherent domain, but it
   is clear that, ideally, such a stick-slip like process generates
   only a high frequency peak at $f=T_0^{-1}$ (and the associated harmonics), and a sharp peak centered at $f=0$ containing the power of the process.
This should occur only if the same critical current (critical angle) was reproduced
at each instability. In reality, the disorder effect causes the distribution of critical currents at the sample scale and necessary induces an irreproducibility
of the elementary instabilities magnitude. This creates
fluctuations around $\langle i_c \rangle$ and a broadening of the
central peak (the low frequency broad band noise). We will
   see that changing the disorder, i.e. the surface roughness, by the reinforcement of the low wave-vectors, results in a change
   of the associated low frequency spectral shape.

\subsection{On the noise for the rough surfaces}

 We have shown that the substantial rise of the critical
 current in the mixed state can be mainly ascribed to the
 increase of the surface roughness. For the samples with the homogeneous surface states, the noise regime is similar above and below
 $B_{c2}$, with comparable values of $C$. For $B \gtrsim
 B_{c2}$, the noise magnitude of the sample (3) (rough surfaces) is enhanced in the same proportion as for the critical
 current, meaning that the same $C$ value is found again. For $B \lesssim B_{c2}$, a noise rise of two orders of magnitude more than $I_c$ is observed.
 It is clear that no fundamental change in the pinning mechanism has been introduced by the
 irradiation,  since this latter was accounted for by the increase of the surface
  roughness.
  In addition, specific heat measurements confirm that the bulk second critical field $B_{c2}$ remains unchanged. Any modifications of
  $\rho_{ff}$ were observed neither. As expected by the very low kinetic energy of the ions, the surface irradiation has no
  influence
  on the bulk properties of the sample. An extra noise regime due to a strong increase of the bulk defects concentration is thus very unlikely.
  An increase of the correlation length $C$ of the surface currents fluctuations, resulting in less statistical averaging, is a more consistent track.
  All samples being identical, apart from their surface state, this increase has to be linked to the introduction of the artificial superficial
  disorder.
  In addition to the enlargement of the coherent domain, a striking change can be observed in the
  spectral shape. The usually smooth decay of the noise power with the frequency is here replaced
  by a strong decrease at high frequencies ($S_{vv} \propto f^{-4.4}$) and a nearly white
  noise below a kink at a frequency $f_c$ (fig. 7). $f_c$ corresponds to a characteristic time $\tau_c = 1/f_c$, above which correlation vanishes. $f_c$ turns out to have a too small
vortex velocity dependence to be obviously connected with a time
of flight or a transit time.
 Anyway, the appearance of a large $C$ and of
 $f_c$ are simultaneous. Since the characteristic length $r_c \approx V_L / f_c$
 is very far away from  $W$ or $a_0$, but always found to be of the order of magnitude of $C$, this suggests that
 $\tau_c$ is associated to the collective rearrangement of the
 surface currents within the coherent domain \cite{note}.

Some large
scale defects can be easily evidenced on the degraded surface (fig.2). To be more precise, we have
plotted the roughness spectrum $S_{\theta\theta}(K)$ in the fig.8,
where the increase of
 spectral weight due to the irradiation is clearly shown.
The lowest wave-vectors are the most affected. They correspond to
the large bumps of diameter $d \gtrsim 1 \mu m$ which are also
visible in the AFM pictures (fig.2). Since they were introduced by
the irradiation, they offer an attractive explanation for the
large $C$ values. We propose that they determine the maximum
available correlation length, within which the instabilities are
correlated, instead of $\delta$ in the virgin samples. This
implies also that a slower diffusion of the instabilities is involved. For such a
large scale mechanism to be effective, the possibility of strong collective effects between vortex endings is necessary. Because of the magnetic long-range interactions, the FLL is much softer for short wave vectors of distortion
(the elastic non-locality) \cite{brandt}.
Even if we do not have a quantitative analysis of the process, the following results are qualitatively accounted for by the introduction of the long-range interactions: (i) $C$ increases
with the magnetic field close to $B_{C2}$ (fig.9), as expected
from the softening of the vortex lattice, and (ii) $C$ returns to the
virgin sample value as soon as $B \gtrsim B_{c2}$. Indeed, the
surface superconducting state is populated by very short vortices
which melt away rapidly in the bulk \cite{kulik}, so their
response is expected to be individual (local). In the absence of
long range interaction in the lattice, the instabilities process
can not involve distortions at small wave-vector, but is limited to
the small range roughness, like in the virgin samples. One of the
key points is that the large scale roughness affects the spatial
extension of each fluctuation, but not their power. The latter is
fixed by the critical current $I_c$ that exhibits a more monotonic
trend towards $B_{c2}$.

It can be observed that $C$ returns strictly to its `` virgin
surface '' value for $B= B^*\approx 1.1 B_{c2}$ and not for $B=
B_{c2}$, where $B_{c2}$ is the bulk value, as measured by specific heat. As previously noted, the surface irradiation
causes a concentration of Ar ions only in the first nm under the
surface. As a consequence, the coherence length is modified by
mean free path effects (for example \cite{ulmaier}) and the
surface second critical field is locally increased, explaining why the surface
properties are now slightly shifted from the bulk properties.
Finally, one can note that any change in the noise statistics can be
observed in any sample and in any regime. The noise is always
Gaussian and stationary, even after the surface irradiation where
the correlation lengths are found to be notably increased. This can be
attributed to the large size of our samples which impedes any
deviation from gaussianity because of statistical averaging. If
the size of the sample is made small enough, a more peculiar
behavior can be expected and is observed  \cite{levy}.

In conclusion, we have studied the critical current and the
voltage noise in Niobium slabs, in the high field regime of the
mixed state and the surface superconducting state. It appears that
the critical current is quantitatively controlled by the surface
roughness, and as a consequence can be modified by controlled low energy irradiation. The mechanism of voltage noise
which reflects the underlying surface current fluctuations is not
modified when the bulk vortex lattice changes into the surface
vortices. The associated correlation length can be associated with the skin length
 of the superficial instabilities. An
artificial surface corrugation with large scale defects causes an enlargement
of this correlation length, only for $B \lesssim B_{c2}$. This reveals likely the long range interactions of the mixed
state .

Acknowledgments: We would like to thank L.~M$\acute{e}$chin
(GREYC, Caen) for the surfaces irradiation, V.~Hardy (CRISMAT,
Caen) for the specific heat measurements, and Bernard Pla\c{c}ais
(ENS, Paris) for fruitful discussion. This work was supported by
la '' r$\acute{e}$gion Basse-Normandie ''.


\newpage

\begin{figure}[tbp]
\centering \includegraphics*[width=10cm]{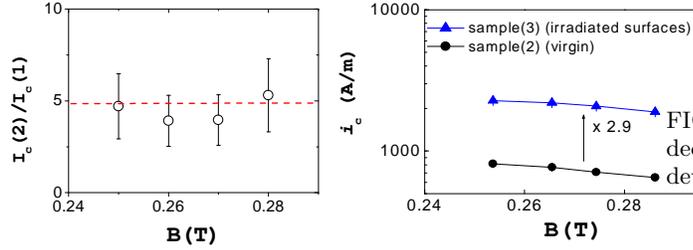} \caption{Color online. Left:
variation of the ratio of the samples (1) and (2) critical
currents, versus the magnetic field (mixed state, $B < B_{c2} =
0.295 T$). The dotted line corresponds to the ratio of their
widths. Right: the surface critical current for the sample (1) and
for the sample (3) (irradiated surfaces). } \label{fig.1}
\end{figure}

\newpage
\begin{figure}[tbp]
\centering \includegraphics*[width=10cm]{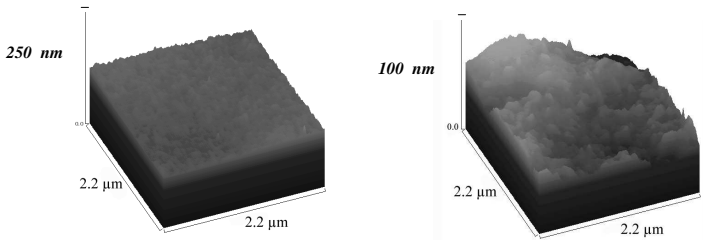} \caption{Left:
AFM picture of the virgin sample (2). Right: AFM picture of the
irradiated sample (3). } \label{fig.2}
\end{figure}

\begin{figure}[tbp]
\centering \includegraphics*[width=6cm]{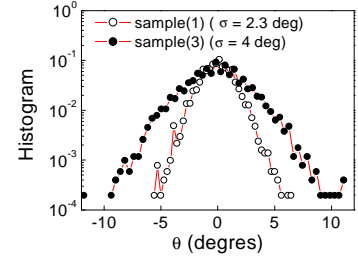}
\caption{Color online. histograms of the roughness angles deduced from the AFM
pictures analysis. The standard-deviations of each histogram are
noted.} \label{fig.3}
\end{figure}

\begin{figure}[tbp]
\centering \includegraphics*[width=6cm]{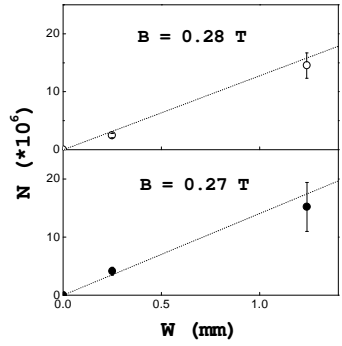} \caption{The
number of fluctuators ($N = S/C^2$), deduced from the noise
values, versus the samples width, for two magnetic fields $B <
B_{C2}$.} \label{fig.4}
\end{figure}

\begin{figure}[tbp]
\centering \includegraphics*[width=6cm]{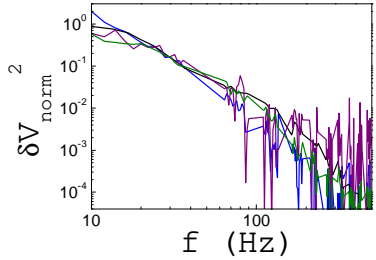}
\caption{Color online. Normalized flux-flow noise spectral densities in
Pb-In (mixed state $B=0.75.B_{c2}, I=3.06 A$ \cite{jo2}), Nb (1)
in the mixed state ($B=0.86.B_{c2}, I=0.8A$) and in the surface
superconducting state ($B=1.12.B_{c2}, I=1A$), Nb (3) (irradiated
surfaces, surface superconducting state only, $B=1.12.B_{c2},
I=4A$) ). $T =$ 4.2 $K$ for all.} \label{fig.5}
\end{figure}

\begin{figure}[tbp]
\centering \includegraphics*[width=6cm]{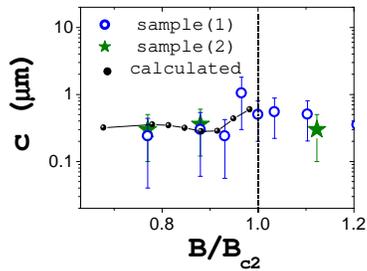} \caption{Color online. The
correlation length in the Nb virgin samples (sample (1) and sample
(2)) as a function of the reduced critical field. Also shown is
the calculated skin depth of the superficial instabilities using
the equation (4) ($T =$ 4.2 $K$).} \label{fig.6}
\end{figure}

\begin{figure}[tbp]
\centering \includegraphics*[width=6cm]{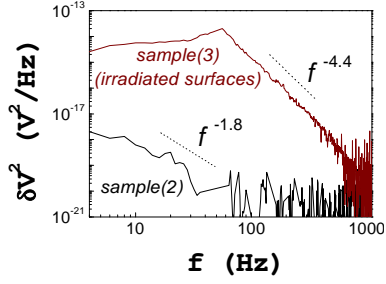}
\caption{Color online. Flux-flow noise spectral density in Nb (1) and (3) ($T =$
4.2 $K$, $B =$ 0.27 $T$ $< B_{c2}$). Note the huge increase of the
noise and the change of the spectral shape after the surface
irradiation.} \label{fig.7}
\end{figure}

\begin{figure}[tbp]
\centering \includegraphics*[width=12cm]{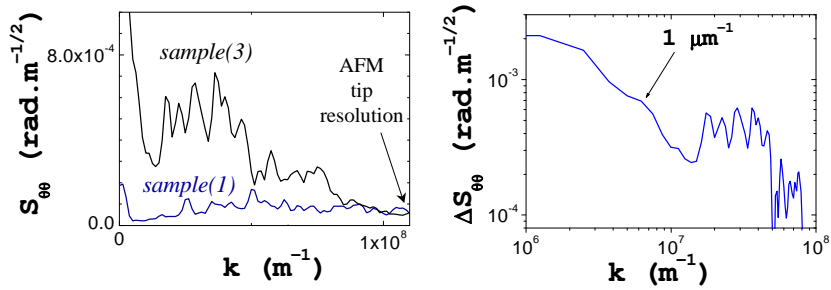} \caption{Color online. Left:
The roughness angle spectral densities $S_{\theta\theta}$ for the
virgin sample (1) and the irradiated sample (3). For this latter,
for small wave-vectors, a strong rise of the spectral density can
be observed. Right: Difference between the two spectral densities
in a log-log scale, so as to evidence more clearly the large
increase of the spectral weight for scales larger than roughly 1
$\mu$m.} \label{fig.8}
\end{figure}

\begin{figure}[tbp]
\centering \includegraphics*[width=6cm]{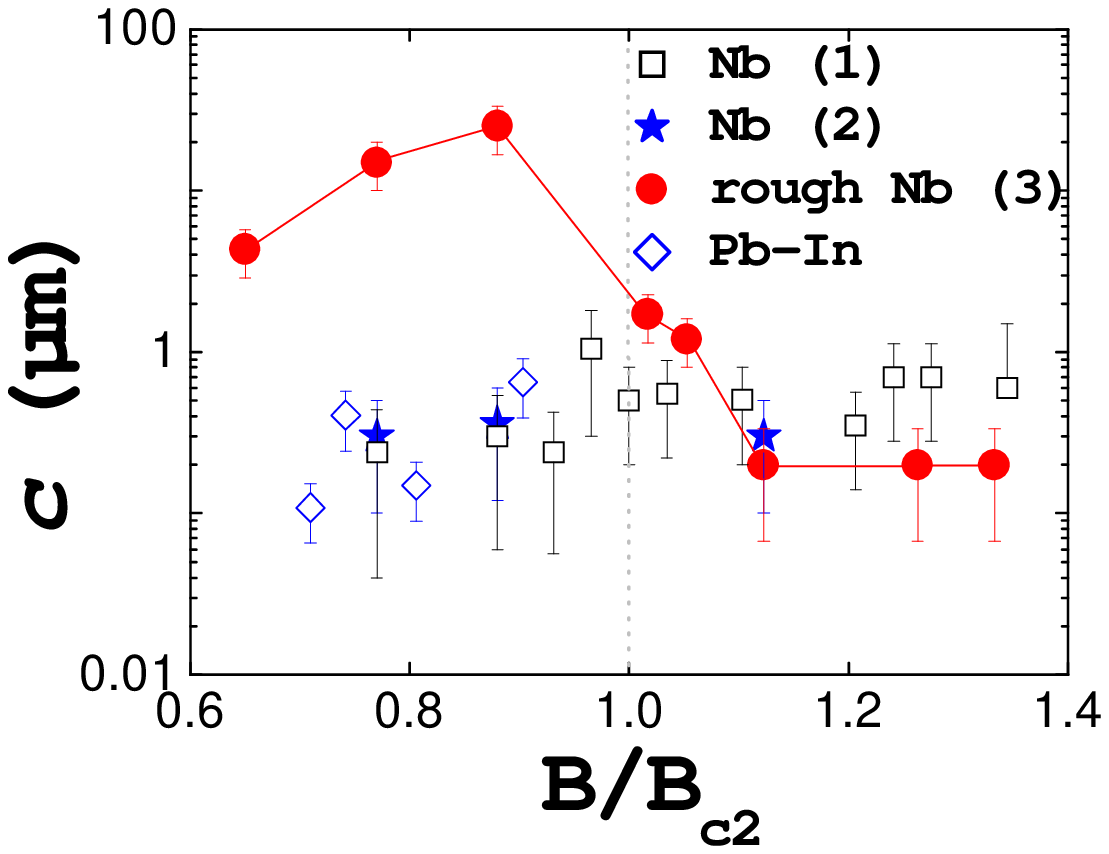} \caption{Color online. The
correlation length as function of the reduced magnetic field
$B/B_{C2}$ ($B_{C2}$ is the bulk value deduced from specific heat
measurements, T= 4.2 K). Note the strong increase of $C$ for $B
\lesssim B^* \approx 1.1 B_{c2}$ for the Nb (3) (irradiated
surfaces).} \label{fig.9}
\end{figure}


\begin{references}
\bibitem{note} We have also observed sharp peaks corresponding to $f_0=V_L/W$  in
 the sample (1). In contrast to $f_c$, $f_0$ can be linked to a time of flight effect. This is an additional `` noise '' mechanism with long range coherence. Since its power
 turns out to be negligible compared to the broad band noise, it can be ignored for the present
 discussion.
\end{references}
\end{document}